\documentstyle[aps,prl,multicol,epsfig]{revtex}

\begin{document}

\draft

\title{Isoscalar Giant Dipole Resonance and Nuclear Matter
       Incompressibility Coefficient}

\author{S. Shlomo$^{1,2)}$ and A. I. Sanzhur$^{1,3)}$}
\address{$^{1)}$Cyclotron Institute, Texas A\&M University,
College Station, TX 77843-3366}
\address{$^{2)}$RIBF Project, Institute of Physical and Chemical Research
(RIKEN), 2-1 Hirosawa, Wako, Saitama 351-0198, Japan}
\address{$^{3)}$Institute for Nuclear Research, Kiev 03028, Ukraine}

\maketitle

\begin{abstract}
We present results of microscopic calculations of the strength function, 
$S(E)$, and $\alpha$-particle excitation cross sections $\sigma(E)$ for 
the isoscalar giant dipole resonance (ISGDR). An accurate and a general 
method to eliminate the contributions of spurious state mixing is presented 
and used in the calculations. Our results provide a resolution to the long
standing problem that the nuclear matter incompressibility coefficient,
$K$, deduced from $\sigma(E)$ data for the ISGDR is significantly smaller 
than that deduced from data for the isoscalar giant monopole resonance
(ISGMR). 
\end{abstract}

\begin{multicols}{2}

Studies of compression modes of nuclei are of particular
interest since their strength distributions, $S(E)$, are sensitive
to the value of the nuclear matter incompressibility coefficient, $K$
\cite{Bohr75,Stringari82}. Over the last two decades, a significant amount 
of experimental work was carried out to identify strength distributions of
the isoscalar giant monopole resonance (ISGMR) in nuclei \cite{Shlomo93}. 
At present, Hartree-Fock (HF) based random-phase-approximation (RPA) 
calculations for the ISGMR reproduce the experimental data for effective 
interactions associated with incompressibility $K=210\pm 20$~MeV 
\cite{Blaizot80}.

The study of the isoscalar giant dipole resonance (ISGDR) is very
important since this compression mode provides an independent source
of information on $K$. Early experimental attempts to identify the
ISGDR in $^{208}$Pb resulted with a value of $E_1\sim 21$~MeV for the
centroid energy \cite{Morsch80,Djalali82}. Similar result for $E_1$
in $^{208}$Pb was obtain in recent experiments \cite{Davis97}. Very 
recent and more accurate data on the ISGDR obtained for a wide range 
of nuclei \cite{Clark99} seems to indicate that the experimental 
values for $E_1$ are smaller than the corresponding HF-RPA results 
by 3 -- 5~MeV.

It was first pointed out in Ref. \cite{Dumitrescu83} that corresponding 
HF-RPA results for $E_1$, obtained with interactions adjusted to reproduce 
the ISGMR data, are higher than the experimental value by more than 3~MeV 
and thus this discrepancy between theory and experiment raises doubts 
concerning the unambiguous extraction of $K$ from energies of compression 
modes (see also Ref. \cite{Colo00}). 

In this work we address this discrepancy between theory and experiment
by examining the relation between $S(E)$ and the excitation cross
section $\sigma(E)$ of the ISGDR, obtained by $\alpha$-scattering. We 
emphasize that it is quite common in theoretical work on giant resonance 
to calculate $S(E)$ for a certain scattering operator $F$ whereas in the 
analysis of experimental data of $\sigma(E)$ one carries out 
distorted-wave-Born-approximation (DWBA) calculations with a certain
transition potential. Here we present results of accurate microscopic 
calculations for $S(E)$ and for $\sigma(E)$ with the folding model (FM) 
DWBA with transition densities $\rho_t(\bbox{r})$ obtained from HF-RPA 
calculations. We provide a simple explanation for the discrepancy 
between theory and experiment concerning the ISGDR and suggest further 
experiments.

We also present an accurate projection method to eliminate the 
contributions of the spurious state mixing (SSM) to $S(E)$ and 
$\rho_t(\bbox{r})$ of the ISGDR obtained in HF-RPA calculations for an 
operator $F$. The method, which is based on the replacement 
of $F$ with a properly modified operator $F_{\eta}$ in the 
calculation of $S(E)$ and $\rho_t(\bbox{r})$, is quite general and 
applicable for any $F$ and for any numerical method used in carrying out 
the RPA calculation, such as configuration space RPA, coordinate space 
(continuum and discretized) RPA and with and without the addition of 
smearing. We note that this method was recently used in the continuum 
RPA calculation of $S(E)$ in Ref. \cite{Gorelik00}. Some preliminary 
results of our calculation were given in Ref. \cite{Kolomiets99}.

In self-consistent HF-RPA calculation one starts by adopting specific 
effective nucleon-nucleon interaction, $V_{12}$, carries out the HF 
calculation for the ground state of the nucleus and then solves the RPA 
equation using the particle-hole (p-h) interaction $V_{ph}$ which 
corresponds to $V_{12}$. The RPA Green's function $G$ 
\cite{Bertsch75,Shlomo75} is obtained from
\begin{equation}
G = G_0 (1+ V_{ph}G_0)^{-1}\ ,
\end{equation}
where $G_0$ is the free p-h Green's function. For
\begin{equation}
F=\sum\limits_{i=1}^{A} f(\bbox{r}_i)\ ,
\label{equ:scop}
\end{equation}
the strength function and transition density are given by
\begin{equation}
S(E)\!=\!\sum\limits_{n}\left|\langle 0 |F| n\rangle\right|^2
\delta(E-E_n)={1\over\pi}\mbox{Im}\left[\mbox{Tr}(fGf)\right],
\label{equ:strace}
\end{equation}
\begin{equation}
\rho_t(\bbox{r},E)={1\over\sqrt{S(E)}}\int f(\bbox{r}\,')
\left[{1\over\pi}\mbox{Im}G(\bbox{r}\,',\bbox{r},E)\right]\,d\bbox{r}\,'\ .
\label{equ:rhot}
\end{equation}
Note that the definition (\ref{equ:rhot}) is consistent with 
\begin{equation}
S(E)=\left|\int\rho_t(\bbox{r},E)f(\bbox{r})\,d\bbox{r}\:\right|^2.
\label{equ:sfun}
\end{equation}

In fully self-consistent HF-RPA calculations, the spurious state
(associated with the center of mass motion) $T=0$, $L=1$ appears at
$E=0$ and no SSM in the ISGDR occurs.
However, although not always stated in the literature, actual
implementations of HF-RPA are not fully self-consistent. One usually
makes the following approximations: (i) neglecting the two-body
Coulomb and spin-orbit interactions in $V_{ph}$, (ii) approximating
momentum parts in $V_{ph}$, (iii) limiting the p-h space
in a discretized calculation by a cut-off energy $E_{ph}^{max}$,
(iv) introducing a smearing parameter (i.e., a Lorenzian with $\Gamma/2$) 
and (v) some numerical approximations. Although the effect of these
approximations on the centroid energies of giant resonances is
small (less than 1~MeV), the effect on the ISGDR is quite serious
since each of these approximations introduces a SSM in the ISGDR.

The first serious attempt to correct for the effect of the SSM on $S(E)$ 
and $\rho_t$, associated with 
$f(\bbox{r})=f_3(\bbox{r})=r^3Y_{1M}(\Omega)$, was presented 
in Ref. \cite{Hamamoto98}. However, this method is not accurate and 
leads to strong reduction in the ISGDR strength at low energies.  In 
Refs. \cite{Deal73,VanGiai81,Dumitrescu83}, the effect of the SSM on 
S(E) was ignored and was only considered with regard to the energy 
weighted sum rule (EWSR) and the derivation of the collective transition 
density.  In a very recent configuration space RPA calculation for the ISGDR
\cite{Colo00} a method for accounting for the SSM was implemented for 
$f_3(\bbox{r})$. However, the results for $S(E)$ and $\rho_t$ depends on 
how the smearing is implemented (not described in the paper).

Let us consider scattering operators, Eq.  (\ref{equ:scop}), with
\begin{equation}
f(\bbox{r})=f(r)Y_{1M}(\Omega)\ ,\ \ \ f_1(\bbox{r})=rY_{1M}(\Omega)\ ,
\end{equation}
and write ${\displaystyle{1\over\pi}}$Im$G$ as sum of separable terms
\begin{equation}
R(\bbox{r}\,',\bbox{r},\omega)=
{1\over\pi}\mbox{Im}G(\bbox{r}\,',\bbox{r},\omega)=
\sum\limits_{a}\rho_a(\bbox{r})\rho_a(\bbox{r}\,')\ .
\label{equ:sepr}
\end{equation}
Note that $\rho_a$ includes the energy dependence and is written as
\begin{equation}
\rho_a(\bbox{r})=\rho_{a3}(\bbox{r})+\rho_{a1}(\bbox{r})\ ,
\label{equ:dssm}
\end{equation}
where $\rho_{a1}(\bbox{r})$ is due to SSM and $\rho_{a3}$, associated
with the ISGDR, fulfills the center of mass condition (for all $a$)
\begin{equation}
\langle f_1\rho_{a3}\rangle=
\int f_1(\bbox{r})\rho_{a3}(\bbox{r})\,d\bbox{r}=0\ .
\label{equ:cond}
\end{equation} 
>From (\ref{equ:sepr}) and (\ref{equ:dssm}) we have in an obvious notation
\begin{equation}
R=R_{33}+R_{31}+R_{13}+R_{11}\ ,
\label{equ:resp}
\end{equation}
with $R_{ij}=\sum\rho_{ai}(\bbox{r})\rho_{aj}(\bbox{r}\,')$,
and the required $S(E)$ and $\rho_t$ can be obtained from
$R_{33}$ using (\ref{equ:rhot}) and (\ref{equ:sfun}) with $f$. 
Since $R_{33}$ is not known we look for a projection operator that
projects out $\rho_{a1}(\bbox{r})$,
\begin{equation}
F_{\eta}=\sum\limits_{i=1}^{A} f_{\eta}(\bbox{r}_i) = F-\eta F_1\ ,
\end{equation}
with $f_{\eta} = f-\eta f_1$. Using (\ref{equ:cond}) and (\ref{equ:resp}) 
we have
\begin{equation}
S_{\eta}(E)=\langle f_{\eta}Rf_{\eta}\rangle=\langle fR_{33}f\rangle+
2\langle fR_{31}f_{\eta}\rangle+\langle f_{\eta}R_{11}f_{\eta}\rangle .
\label{equ:seta}
\end{equation}
Note that $\langle f_{\eta}R_{11}f_{\eta}\rangle$ is minimum for
\begin{equation}
\eta=\langle fR_{11}f_1\rangle/\langle f_1 R_{11}f_1\rangle=
\sum\langle f\rho_{a1}\rangle\langle f_1\rho_{a1}\rangle/
\sum\langle f_1\rho_{a1}\rangle^2\ ,
\end{equation}
and for the last two terms in (\ref{equ:seta}) to vanish we must have
\begin{equation}
\langle f\rho_{a1}\rangle=\eta\langle f_1\rho_{a1}\rangle,\ \ \ \,
\mbox{for all}\ a.
\label{equ:frho}
\end{equation}
The condition (\ref{equ:frho}) holds in case (\ref{equ:sepr}) has only
one term, such as in a discretized calculation without smearing or in
a configuration space calculation, and/or in case all
$\rho_{a1}(\bbox{r})$ are proportional to the same transition density,
the coherent spurious states transition density \cite{Bertsch83}
\begin{equation}
\rho_{a1}(\bbox{r})=\alpha_a\rho_{ss}(\bbox{r})=
\alpha_a{\partial\rho_0\over\partial r}\,Y_{1M}(\Omega)\ ,
\label{equ:spur}
\end{equation}
where $\rho_0$ is the ground state density of the nucleus. The value of 
$\eta$ associated with $\rho_{ss}$ is then given by
\begin{equation}
\eta=\langle f\rho_{ss}\rangle/\langle f_1\rho_{ss}\rangle.
\label{equ:eta}
\end{equation}

To determine $\rho_t$ for the ISGDR we first use (\ref{equ:rhot}),
(\ref{equ:sepr}), (\ref{equ:dssm}), (\ref{equ:cond}) and (\ref{equ:frho}) 
with $F_{\eta}$ and obtain
\begin{equation}
\rho_{\eta}(\bbox{r})={1\over\sqrt{S_{\eta}(E)}}
\sum c_a[\rho_{a3}(\bbox{r})+\rho_{a1}(\bbox{r})]\ ,
\label{equ:tden}
\end{equation}
with $c_a=\langle f_{\eta}\rho_{a3}\rangle$.  To project out the spurious
term from (\ref{equ:tden}) we make use of (\ref{equ:cond}) with
$\rho_{a1}=\alpha_a\rho_{ss}$ and obtained
\begin{equation}
\rho_t(\bbox{r})=\rho_{\eta}(\bbox{r})-\alpha\rho_{ss}\ ,\ \ \ 
\alpha=\langle f_1\rho_{\eta}\rangle/\langle f_1\rho_{ss}\rangle\ .
\label{equ:alph}
\end{equation}
It is important to emphasize that using $f_{\eta}$ in (\ref{equ:sfun})
with $\rho_t(\bbox{r})$ from (\ref{equ:alph}), one obtains the required
$S_{\eta}(E)$. This is due to the fact that the averaging process in
(\ref{equ:tden}) was carried out using $F_{\eta}$ 
and not $F$. Use of $F$ in (\ref{equ:rhot}) may produce erroneous results 
for $\rho_t(\bbox{r})$ in (\ref{equ:alph}) in case there are several terms 
in (\ref{equ:sepr}). 

We now limit our discussion to the operator
$F_3=\sum\limits_{i=1}^{A} f_3(\bbox{r}_i)$, where we have for 
(\ref{equ:spur})
\begin{equation}
\eta={5\over 3}\langle r^2\rangle\ .
\label{equ:eta3}
\end{equation}
We note that the values of $\eta$ obtained for $\rho_a$ associated with 
single p-h transitions in the 1$\hbar\omega$ region and the RPA results 
for the spurious state $\rho_t$, differ from that of (\ref{equ:eta3}) 
by less than 20\% \cite{Shlomo00}.

We have carried out numerical calculations for the $S(E)$, $\rho_t(\bbox{r})$ 
and $\sigma(E)$ within the FM-DWBA-HF-RPA theory. We used the SL1 Skyrme 
interaction \cite{Liu91}, which is associated with $K=230$~MeV, and carried 
out HF calculations using a spherical box of $R\geq 25$~fm.  For the RPA 
calculations we used the Green's function approach with mesh size 
$\Delta r=0.3$~fm and p-h maximum energy of $E_{ph}^{max}=150$~MeV 
(we include particle states with principle quantum number up to 12), since it
is well-known that in order to extract accurate $\rho_t(\bbox{r})$,
$E_{ph}^{max}$ should be much larger than the value required
($E_{ph}^{max}\sim 50$~MeV) to recover EWSR. Since in our calculation 
we also neglected the two-body coulomb and spin-orbit interactions, the
spurious state energies differ from 0 by a few MeV. We therefore renormalized
the strength of the $V_{ph}$ by a factor (0.99 and 0.974 for $^{116}$Sn and 
$^{208}$Pb, respectively), to place the spurious state at E=0.2 MeV. We have 
included a Lorenzian smearing ($\Gamma/2=1$~MeV) and corrected for the 
SSM as described above. We carried out the FM-DWBA calculation for 
$\sigma(E)$ using a density dependent Gaussian nucleon-$\alpha$ interaction 
with parameters adjusted to reproduce the elastic cross section, with 
$\rho_0$ and $\rho_t$ from HF-RPA, see Ref. \cite{Kolomiets00}.

Using the operator $f=r^2$ for the ISGMR we calculated the 
corresponding $S(E)$, for $E$ up to 60 MeV. We 
recover $100\%$ of the corresponding EWSR and obtained the values of 17.09 
and 14.48 MeV for the centroid energy of the ISGMR in $^{116}$Sn and 
$^{208}$Pb, respectively. The corresponding recent experimental values are 
$16.07 \pm 0.12$ and $14.17 \pm 0.28$ MeV, respectively \cite{Youngblood99}.

In Figure 1, we present the results of the $ES_3(E)$ and $ES_1(E)$  
for $^{116}$Sn, obtained by using Eq. (\ref{equ:strace}) with $f_3$ and 
$f_1$, respectively. Note that from (\ref{equ:cond}) and (\ref{equ:resp}) 
$S_1(E)=\langle f_1 R_{11}f_1\rangle=\sum\langle f_1\rho_{a1}\rangle^2$\ ,
provides a measure to the contribution of the SSM to $S_3(E)$. The large 
contribution from the Lorenzian tail of the spurious state is clearly seen
in the figure. Also shown in Figure 1, is the ratio 
$\langle f_3Rf_1\rangle/\langle f_1 Rf_1\rangle$. At low energy
this ratio is very close to $\eta$, reflecting the fact that
the transition density at the spurious state energy is close to that of Eq.  
(\ref{equ:spur}) \cite{Shlomo00}. At higher energies, this ratio exhibits 
fluctuations due to the nonnegligible terms $R_{31}$, Eq. (\ref{equ:resp}).

Figure 2, exhibits the strength functions for the ISGDR in $^{208}$Pb 
obtained from Eqs. (\ref{equ:sfun}), (\ref{equ:rhot}) and (\ref{equ:alph}). 
The solid line describes the result obtained using $f_{\eta}$. Note that 
this result coincides with $S_{\eta}(E)$, which is free of SSM
contribution. Similarly, the dashed line describes the erroneous result 
obtained using $f_3$ (it is also different from $S_3(E)$). We find that
when using $f_3$, the excitation strengths obtained for certain states 
are sensitive to the value of $\Gamma$. The result obtained with $f_3$
coincides with that obtained with $f_{\eta}$ for $\Gamma\rightarrow$ 0,
as expected since in this case Eq. (\ref{equ:sepr}) has one term. Thus,
in configuration RPA calculation of $\rho_t$, one may use $f_3$ and 
correct for the SSM contribution before the smearing process.

\vspace*{0.1in}
\centerline{\epsfxsize=3.3in\epsffile{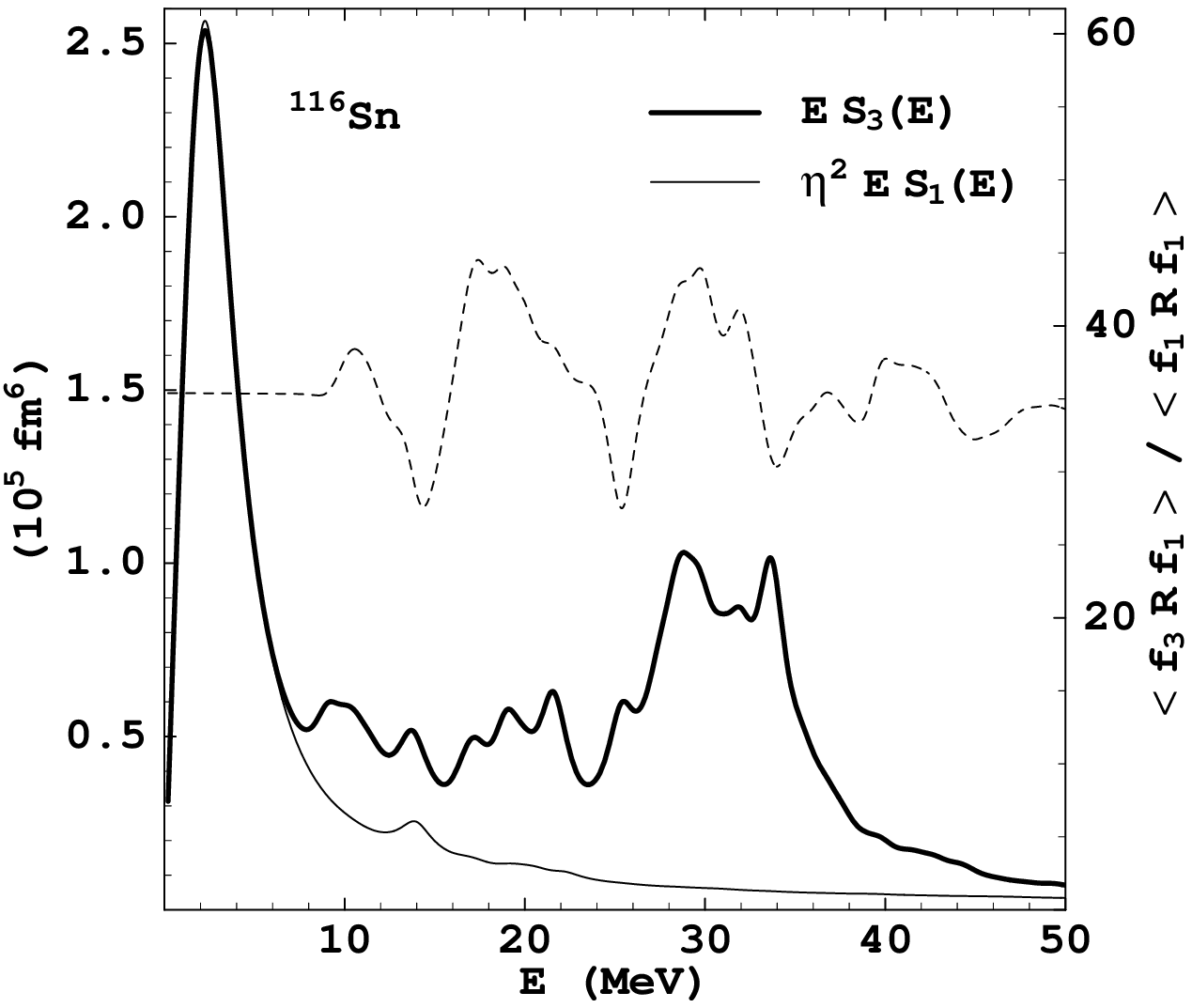}}
\vspace*{0in}
{\small FIG. 1.
Energy weighted strength functions for ISGDR in $^{116}$Sn, obtained from 
Eq. (\ref{equ:strace}) for the scattering operators $f_3$ (thick line) 
and $f_1$ (thin line) with $\eta={5\over 3}\langle r^2\rangle=35.6$ fm$^2$. 
Also shown is the ratio 
$\langle f_3Rf_1\rangle/\langle f_1 Rf_1\rangle$ (dashed line).}
\vspace{0.2in}

\centerline{\epsfxsize=3.3in\epsffile{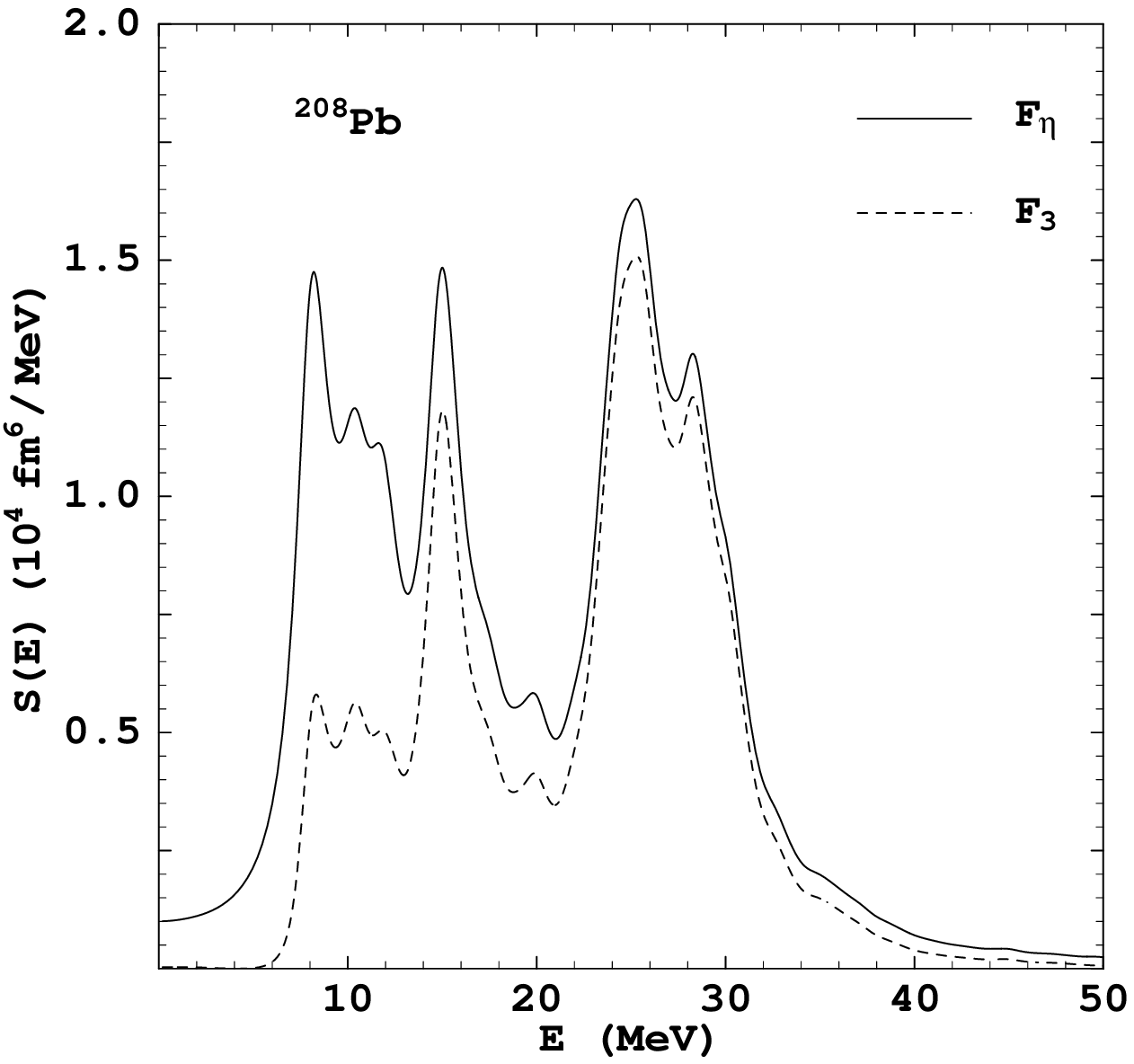}}
{\small FIG. 2.
Strength functions for the ISGDR in $^{208}$Pb obtained from Eqs. 
(\ref{equ:rhot}), (\ref{equ:alph}) and (\ref{equ:sfun}), using $f_3$ 
(dashed line) and $f_{\eta}=f_3-\eta f_1$ (solid line), with 
$\eta=52.1$ fm$^2$.}
\vspace*{0.2in}

Our results for the ISGDR, $S_{\eta}(E)$, indicate two main components 
with the low energy component containing close to 30\% of the EWSR (for 
$E$ up to 23 and 19 MeV for $^{116}$Sn and $^{208}$Pb, respectively), in 
agreement with the experimental observation \cite{Clark00}. Similar 
results were obtained for other nuclei and in other calculations 
\cite{Kolomiets99,Shlomo00,Colo00}

In Figure 3 we present results of microscopic calculations of the 
excitation cross section of the ISGDR in $^{116}$Sn by 240 MeV 
$\alpha$-particle, carried out within the FM-DWBA. The dashed lines are 
obtained using $\rho_{coll}(r)$ of the ISGDR \cite{Stringari82,VanGiai81}.
It is seen from the upper panel that the use of $\rho_{coll}$ increases
the EWSR by at least 10\% and

\vspace*{0in}
\centerline{\epsfxsize=3.3in\epsffile{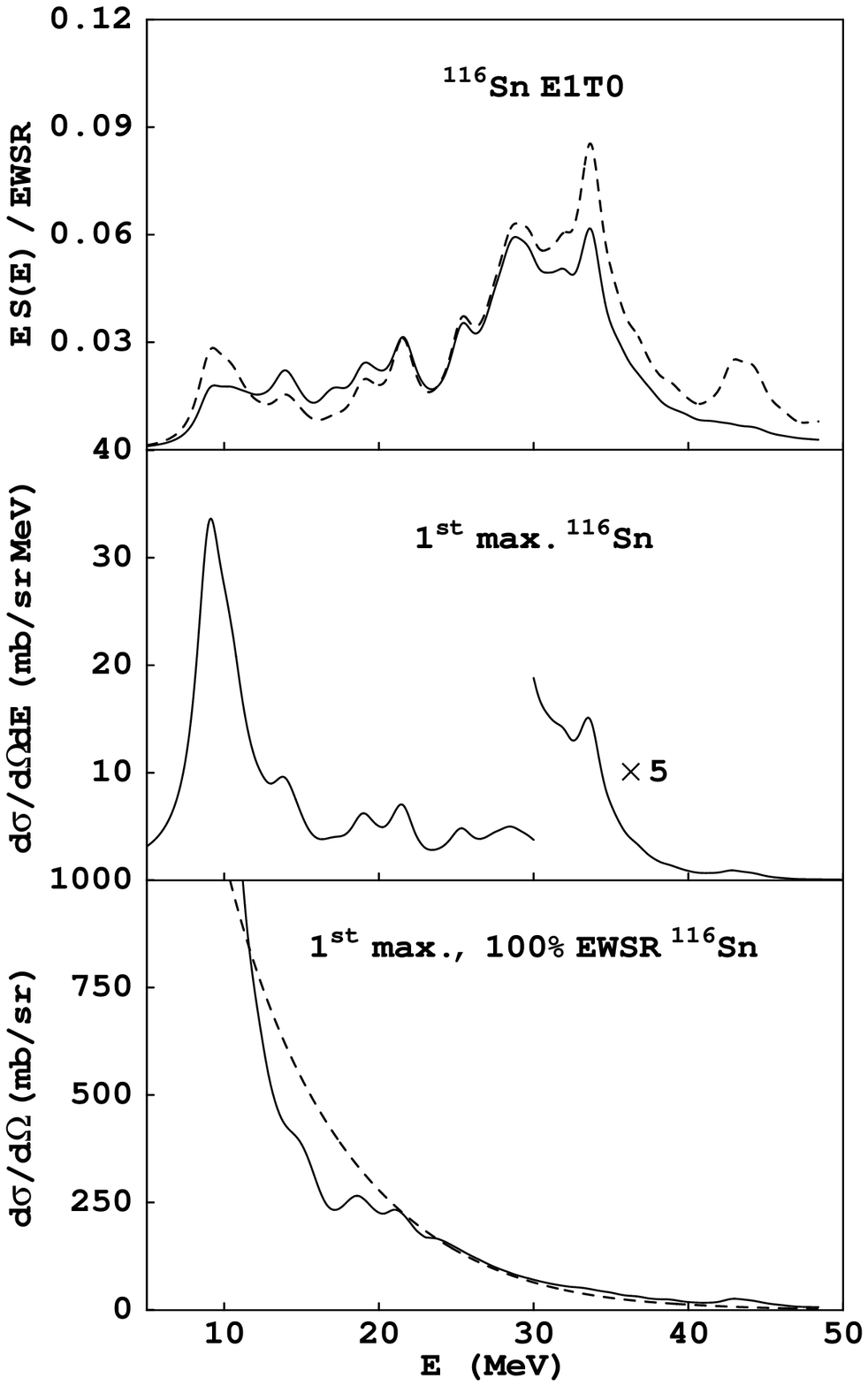}}
\vspace*{-0.1in}
{\small FIG. 3.
The ISGDR in $^{116}$Sn. The middle panel: maximun double differential cross 
section obtained from $\rho_t$ (RPA). The lower panel: maximun cross section 
obtained with $\rho_{coll}$ (dashed line) and $\rho_t$ (solid line) 
normalised to 100\% of the EWSR. Upper panel: The solid and dashed 
lines are the ratios of the middle panel curve with the solid and dashed 
lines of the lower panel, respectively.}  
\vspace*{0.1in}

\noindent
may shift the centroid energy by a
few percent. An important result of our calculation is that the
maximum cross section for the ISGDR drops below the current experimental
sensitivity of 2 mb/sr/MeV for excitation energy above 35 MeV (30 MeV 
for $^{208}$Pb), which contains about 20\% of the EWSR. This missing
strength leads to a reduction of more than 2.5 MeV in the ISGDR energy
and thus explains the discrepancy between theory and experiment. More
sensitive experiments and/or with higher $\alpha$-particle energy are
thus needed.

In summary, we described and applied an accurate and general method to 
eliminate the SSM contributions from $S(E)$ and 
$\rho_t$. Our results indicate: (i) Existence of non-negligible ISGDR 
strength at low energy and (ii) Accurate determination of the relation 
between $S(E)$ and $\sigma(E)$ resolves the long standing problem of
the conflicting results obtained for $K$, deduced from experimental 
data $\sigma(E)$ for the ISGDR and data for the ISGMR.

We thank Professors A. Arima and I. Hamamoto for interesting discussions.
This work was supported in part by the US Department of Energy under 
grant no. DOE-FG03-93ER40773.

\end{multicols}

\begin{references}

\bibitem{Bohr75} A.~Bohr and B.~Mottelson, {\em Nuclear Structure} 
(W. A. Benjamin, London, 1975), Vol. II, Chap. 6.

\bibitem{Stringari82} S.~Stringari, Phys. Lett. {\bf 108B}, 232 (1982).

\bibitem{Shlomo93} S.~Shlomo and D.~H.~Youngblood, Phys. Rev. C {\bf 47}, 
529 (1993).

\bibitem{Blaizot80} J.~P.~Blaizot, Phys. Rep. {\bf 64}, 171 (1980).

\bibitem{Morsch80} H.~P.~Morsch, M.~Rogge, P.~Turek, and C.~Mayer-Boricke,
Phys. Rev. Lett. {\bf 45}, 337 (1980).

\bibitem{Djalali82} C.~Djalali, N.~Marty, M.~Morlet, and A.~Willis,
Nucl. Phys. {\bf A380}, 42 (1982).

\bibitem{Davis97} B.~Davis {\em et al.}, Phys. Rev. Lett. {\bf 79}, 609 
(1997).

\bibitem{Clark99} H.~L.~Clark, Y.-W.~Lui, D.~H.~Youngblood, K.~Bachtr,
U.~Garg, M.~N.~Harakeh, and N.~Kalantar-Nayestanski, Nucl. Phys. {\bf A649}, 
57c (1999).

\bibitem{Dumitrescu83} T.~S.~Dumitrescu, and F.~E.~Serr, Phys. Rev. 
C {\bf 27}, 811 (1983).

\bibitem{Colo00} G.~Colo, N.~Van~Giai, P.~F.~Bortignon and M.~R.~Quaglia
Phys. Lett. {\bf B485}, 362 (2000).

\bibitem{Gorelik00} M.~L.~Gorelik, S.~Shlomo and M.~H.~Urin, Phys. Rev. 
C {\bf 62}, 044301 (2000).

\bibitem{Kolomiets99} A.~Kolomiets, O.~Pochivalov, and S.~Shlomo,
Progress in Research, Cyclotron Institute, Texas A\&M University,
April 1, 1998 - March 31, 1999,  III-1 (1999).

\bibitem{Bertsch75} G.~F.~Bertsch and S.~F.~Tsai, Phys. Rep. {\bf 18}, 
125 (1975).

\bibitem{Shlomo75} S.~Shlomo and G.~F.~Bertsch, Nucl. Phys. {\bf A243}, 
507 (1975).

\bibitem{Hamamoto98} I.~Hamamoto, H.~Sagawa, and X.~Z.~Zhang, Phys. Rev. 
C {\bf 57}, R1064 (1998).

\bibitem{Deal73} T.~J.~Deal, Nucl. Phys. {\bf A217}, 210 (1973).

\bibitem{VanGiai81} N.~Van~Giai and H.~Sagawa, Nucl. Phys. {\bf A371}, 
1 (1981).

\bibitem{Bertsch83} G.~F.~Bertsch, Suppl. Progr. Theor. Phys. {\bf 74}, 
115 (1983).

\bibitem{Shlomo00} S.~Shlomo et al, to be published.

\bibitem{Liu91} K.-F.~Liu, H.-D.~Lou, Z.-Y.~Ma, Q.-B.~Shen, Nucl. Phys. 
{\bf A534}, 1 (1991); {\bf A534}, 25 (1991).

\bibitem{Youngblood99} D.~H.~Youngblood, H.~L.~Clark and Y.-W.~Lui,  
Phys. Rev. Lett. {\bf 82}, 691 (1999).

\bibitem{Clark00} H.~L.~Clark, Y.-W.~Lui and D.~H.~Youngblood,
Progress in Research, Cyclotron Institute, Texas A\&M University, 
April 1, 1998 - March 31, 1999, I-9 (1999);
April 1, 1999 - March 31, 2000, I-17 (2000).

\bibitem{Kolomiets00} A.~Kolomiets, O.~Pochivalov, and S.~Shlomo,
Phys. Rev. C {\bf 61}, 034312 (2000).

\end{references}
\end{document}